\documentclass[10pt]{article}
\usepackage[cp866]{inputenc}
\usepackage[russian]{babel}

\begin{document}


\newcommand{\N}{N\raise.7ex\hbox{\underline{$\circ $}}$\;$}

\begin{center}
{\bf

\thispagestyle{empty}

BELARUS NATIONAL ACADEMY OF SCIENCES

B.I. STEPANOV's INSTITUTE OF PHYSICS

}
\end{center}

\vspace{20mm}

\begin{center}
{\bf
Red'kov V.M.\footnote{E-mail: redkov@dragon.bas-net.by}}

\end{center}

\vspace{10mm}

\begin{center}
{\bf
   ON CONCEPT OF PARITY FOR A FERMION
}

\end{center}

\begin{center}
{\bf Abstract}
\end{center}

\begin{quotation}

The known problem of fermion parity is considered on the base of
investigating possible liner single-valued representations of
spinor coverings of the extended Lorentz group. It is shown that
in the frame of this theory does not exist, as separate concepts,
P-parity and T-parity for a fermion, instead only some unified
concept of (PT)-parity can be determined in a group-theoretical
language.

\end{quotation}

\vspace{5mm}

{\bf PASC: 1130, 0230, 0365, 2110H}

\vspace{5mm}


\vspace{10mm}

8 pages, 17 references

\newpage

{\bf 1. On spinor space-time structure}

It is well-known that the space-time vector  $x^{a}$ = $( t, x, y,z)$
can be identified with the explicit realization of the one-valued simplest
representation of the (restricted) Lorentz group
$L^{\uparrow}_{+}$. But in nature we face particles of integer and half-integer spin,
 bosons and fermions. Among respective sets of Lorentz group representations, entities
 describing bosons and fermions,  there exists a clear-cut distinction: boson-based representations
$T_{bos.}$  are single-valued whereas fermion-based representations $T_{ferm.}$  are double-valued
on the group  $L^{\uparrow}_{+}$. In other words, $T_{bos.}$ are global representations, whereas
 $T_{ferm.}$ are just local ones.
 One might have advanced a number of theoretical arguments to neglect  such a light trouble.
 However, the fact of prime importance is that this global-local difficulty cannot
 be cleared up --
  in the frames  of the orthogonal group $L^{\uparrow}_{+} = SO_{0}(3.1)$ it is   insurmountable
  [1-33].

 It has long been known that to work against the global-local problem one must investigate and employ
  one-valued representations (boson-based as well as fermion-based) of the covering group $SL(2.C)$.
 The group $SL(2.C)$ is other group, different from
$L^{\uparrow}_{+} = SO_{0}(3.1)$, but it is linked to  the latter by a quite definite homomorphic
 mapping [1-5]. At this every local representation of the orthogonal group has its counterpart --
  global representation of the covering group; evidently,  it is a quite usual procedure.

In essence,  a strong form of this changing
$L^{\uparrow}_{+} = SO_{0}(3.1) \;\; \Longrightarrow \;\; SL(2.C)$, when
 instead of the
orthogonal Lorentz group  $L^{\uparrow}_{+} = SO_{0}(3.1)$   we are going
to employ the  covering group
$SL(2.C)$ and its  representations throughout, and in addition we are going to work in the same manner  at describing
 the space-time structure itself, is the known Penrose-Rindler spinor  approach.
  Else one point  should be
 emphasized in this context: there exists  a close connection between the above  spinor approach in physics
 and the known Kustaanheimo-Stiefel technics
 \footnote{The author is grateful to Yu.A. Kurochkin for  pointing out this connection and
 the  desirability  to have investigated it in more detail; also the author is grateful to E.A. Tolkachev
 for explanation of the main ideas of the Kustaanheimo-Stiefel approach.}.
 Even more, one may say that these two methods are intrinsically the same, though
 looking differently.

An essential feature of the present work is the use of  4-spinors (bispinors in physical terms)
instead of 2-spinors that are the base for the most literature.
 The foundation to do so is that among 4-spinors there exists real-valued ones, so-called
  Majorana 4-spinors [6].
The simplest  real-valued 4-spinor depends upon  four real parameters, exactly as in the case of Weyl complex 2-spinor.
But these two models are not equivalent  because all properties under  spinor discrete  symmetry
transformations are immersed in the 4-spinor technics from the very beginning, whereas in the
2-spinor model it is not
so.

Therefore,  starting from real 4-spinors, as fundamental physical objects, one may
expect to embed discrete symmetry
operations in spinor form and simultaneously   to trace  in detail the appearance
of the  imaginary $i$ in the model, which is of prime importance in quantum theory
(for example, by translating the real Majorana wave equation into a complex Dirac's).

The present work is restricted to investigating only the problem of  accurate description
of the one-valued representations of four different spinor groups, all covering the full Lorentz group
$L ^{\uparrow \downarrow} _{+-}$, including $P$ ant $T$- reflections.
The results obtained will provide us with the base
for a new discussion of the old fermion parity problem [9-34].

{\bf 2. Spinor covering of the total Lorentz group  $L_{+-}^{\uparrow \downarrow}$}

By adding to the set of proper orthochronous Lorentz matrices
$$
L^{\;\;b}_{a}(k, k^{*}) = \bar{\delta }^{b}_{c}\;
[\;  - \delta ^{b}_{c} \; k^{n}\; k^{*}_{n} \; + \;
k_{c} \; k^{b*} \; +  \;k^{*}_{c} \;k^{b}  \; +
$$
$$
+ \; i\;
\epsilon ^{\;\;bmn}_{c}\;  k_{m}\; k^{*}_{n}\; ]  \; , \;\;
 L(k, k^{*}) = L(-k, -k^{*})  \;
\eqno(2.1a)
 $$

\noindent two matrices
$$
 P: \;\; L^{(P)b}_{a} = + \bar{\delta}^{b}_{a} \;,  \qquad T:
\;\;  L^{(T)b}_{a} = - \bar{\delta }^{b}_{a} \;
\eqno(2.1b)
 $$

\noindent one readily produces the total Lorentz group $L ^{\uparrow \downarrow} _{+-}$.
The commutation rules between
$L^{\;\;b}_{a}(k, k^{*})$  and  the discrete elements   $P,T$
are defined by the relation
$$
 \bar{\delta}^{b}_{a} \;
L^{\;\;c}_{b}(k, k^{*}) = L^{\;\;b}_{a}(\bar{k}^{*}\; , \bar{k})\;
\bar{\delta }^{c}_{b} \; .
\eqno(2.1c)
$$

\noindent
It is  known that the group $L ^{\uparrow \downarrow} _{+-}$ has four types of vector representations:
 $$
T^{\;\;b}_{a} (L) = f(L) \; L^{\;\;b}_{a} \; , \;\; L \in
L^{\uparrow \downarrow}_{+-}    \; .
\eqno(2.2a)
 $$

\noindent Here  $f(L)$ stands for a real-valued function on the group $L^{\uparrow \downarrow}_{+-}$
obeying the equation
$$
f(L_{1})\; f(L_{2}) = f(L_{1} \; L_{2}) \; .
\eqno(2.2b)
$$

\noindent Eq.  (2.2b) can be satisfied by four different functions  ($L \in  L^{\uparrow \downarrow}_{+-}$):
$$
f_{1}(L) = 1 \; , \; f_{2}(L) = \det (L) \;, \; $$ $$ f_{3} (L) =
sgn \; (L^{\;\;0}_{0})\; ,\;  f_{4}(L) = \det (L)\; sgn \;
(L^{\;0}_{0})\; .
\eqno(2.2c)
 $$

\noindent Correspondingly there exist four 4-vector representations of the  full
Lorentz group :
$$
\left.  \begin{array}{llll}
1. & T_{1}(L) =\;\; L    & 2. &  T_{2}(L)  = \;\;L     \\
   & T_{1}(P) = + P  &    &  T_{2}(P)  = - P   \\
   & T_{1}(T) = + T  &    &  T_{2}(T)  = - T   \\  [2mm]
3. & T_{3}(L) = \;\; L    & 4. &  T_{4}(L) = \;\; L      \\
   & T_{3}(P) = + P  &    &  T_{4}(P) = - P    \\
   & T_{3}(T) = - T  &    &  T_{4}(T) = + T     \end{array} \right. .
\eqno(2.3)
 $$

\noindent It is readily verified that these four are non-equivalent ones.

It should be emphasized that the above-described  enlargement of the group
 $L^{\;\; b}_{a}(k,k^{*})$  by  adding two discrete operations
 $P$   and  $T$ is not an enlargement of the spinor group
 $SL(2.C)$, instead it is only an expansion of the orthogonal group
$L^{\uparrow }_{-}$. From  a spinor view-point  the  operations  $P$    and    $T$
are transformations acting in the space of  2-rank spinors but not in the space of 1-rank spinors.
Often,  all group-theoretical   consideration of discrete symmetry problems is  reduced only
to this treatment. Evidently, a more comprehensive study of $P,T$-symmetry can be done in the framework
of first-rank  spinors, when one extends the covering group $SL(2.C)$ by adding  \underline{spinor} discrete operations.

Now we can start to solve this task.
A covering group for the total Lorentz group can be made  constructively   by
adding two specific  $4\times 4$-matrices to the known set of bispinor transformations  $\tilde{SL}(2.C)$
of the group $SL(2.C)$
$$
S(k,
\bar{k}^{*}) = \left ( \begin{array}{cc} B(k) & 0 \\ 0 &
B(\bar{k}^{*}) \end{array} \right ) \in  \tilde{SL}(2.C) \sim
SL(2.C) .
\eqno(2.4)
$$

\noindent Those two new matrices  are to be taken from the following
$$
M = \left ( \begin{array}{cc} 0 & I  \\
I & 0 \end{array} \right ) \; , \;   \;
M'  = i M \;  , \;\;
N =\left ( \begin{array}{cc} 0 & -i I
\\ + i I & 0 \end{array} \right ) \; ,  \; \;
'N = i N   \;
\eqno(2.5)
 $$

\noindent with the multiplication  table
$$
\left.  \begin{array}{ccccc} &  M   &   M ' &  N  &  'N  \\ [4mm]
M\;\; &  \left ( \begin{array}{cc} \;\;\;\;I & \;\;\;\;0 \\ \;\;\;\;0 & \;\;\;\;I
\end{array} \right ) & \left ( \begin{array}{rr}
\;\;\;iI & \;\;\;\; 0 \\ \;\;\;\; 0 & \;\;\; iI \end{array} \right )   &
\left ( \begin{array}{rr}  +iI & \;\;0 \\ \;\;0 & -iI \end{array} \right ) &
\left ( \begin{array}{rr} \;-I & \;\;0 \\ \;\;0 &  \;+I \end{array} \right ) \\[4mm]
M'\;\;  & \left ( \begin{array}{rr} +iI & \;\; 0 \\ \;\; 0 & +iI \end{array} \right ) &
\left ( \begin{array}{rr} \;\;-I & \;\;\; 0 \\ \;\;\; 0 & \;-I \end{array} \right ) &
\left ( \begin{array}{rr} \;-I & \;\;\; 0 \\ \;\;\;0 & \;+I \end{array} \right ) &
\left ( \begin{array}{rr} -iI &\;\; 0 \\ \;\;0 & +iI \end{array} \right ) \\   [4mm]
N \;\; &  \left ( \begin{array}{cc} -iI & \;\;0 \\ \;\;0 & +iI \end{array} \right )  &
\left ( \begin{array}{rr} \;+I & \;\;\;0 \\ \;\;\;0 & \;-I \end{array} \right ) &
\left ( \begin{array}{rr} \;+I & \;\;\;0 \\ \;\;\;0 & \;+I \end{array} \right ) &
\left ( \begin{array}{rr} +iI & \;\;0 \\ \;\;0 & +iI \end{array} \right )    \\ [4mm]
'N \;\;  &  \left ( \begin{array}{cc} \;+I &\;\;\; 0 \\ \;\;\;0 & \;-I \end{array} \right ) &
\left ( \begin{array}{rr} +iI & \;\;0 \\ \;\;0 & -iI  \end{array} \right ) &
\left ( \begin{array}{rr} +iI  & \;\;0 \\ \;\;0 & +iI \end{array} \right ) &
\left ( \begin{array}{rr} \;-I & \;\;\; 0 \\ \;\;\;0 & \;-I \end{array} \right )
\end{array} \right.
\eqno(2.6)
$$

It should be noted that the squared matrices $M , M', N, 'N$ equal
 $+I,-I \in \tilde{SL}(2.C) $.
  Therefore, having added  any two elements from these  we will have extended the group
 $\tilde{SL}(2.C)$  in fact by two new operations only.
 Also, because the group $L(2.C)$
contains $-I$, extension of the group  by any two from  $-M , -M', -N, -'N$ results in the same.

However, if one chooses any other different from
+ $1 , - 1, + i , - i$  phase factors at $M , M', N, 'N$, then one will obtain
 substantially new extended groups. For instance,
addition of the discrete element $\alpha M$ with  $\alpha = \exp (i2\pi /n) $
leads to a group extended by  $(n - 1)$ new elements:
$$
\alpha \; M \; , \; \alpha ^{2} \; I\;  ,\; \alpha ^{3} \; M\; ,
\alpha ^{4} \; , \; \ldots , \; \alpha ^{(n-1)}\;  M \; .
 $$

\noindent Those will not be considered in the
present work.

So, enlarging the group  $\tilde{SL}(2.C)$ as was shown above, we obtain six covering groups:
$$
\left. \begin{array}{llll}
G_{M}  & =     \{\; S(k,\bar{k}^{*}) \uplus  M \uplus  M' \; \}   \; ,     &
\qquad
G_{N}  & =   \{\; S(k,\bar{k}^{*}) \uplus  N \uplus \; 'N \;\}  \; ,     \\
G'     & =   \{ \; S(k,\bar{k}^{*}) \uplus M'  \uplus N \; \}   \; ,     &
\qquad
 'G    & =   \{ \;  S(k,\bar{k}^{*}) \uplus 'N \uplus M \; \}   \; ,     \\
G      & =   \{ \; S(k,\bar{k}^{*}) \uplus  M \uplus N \; \}    \; ,  &
\qquad
'G'    &  =  \{ \; S(k,\bar{k}^{*}) \uplus M'  \uplus 'N \;\} \;
\end{array} \right.
\eqno(2.7)
$$

\noindent with  multiplication tables
$$
\left. \begin{array}{lrrr}
G_{M}: & M^{2} = + I \; ,\;\; &  M^{'2} = - I \; , & \;\; M M'  = M'  M \; ; \\
G_{N}: & N^{2} = + I \; ,\;\; & 'N^{2} = - I  \; , & \;\; N  'N = 'N N  \; ; \\
G' :   & M^{'2}= - I \; ,\;\; &  N^{2} = + I  \; , & \;\; M' N = - N M' \; ; \\
'G :   & 'N^{2} = - I\; ,\;\; &  M^{2} = + I  \; , & \;\; 'N M = - M 'N \; ;  \\
G  :   & M^{2} = +  I\; ,\;\; &  N^{2} = + I  \; , & \;\; M N = - N M   \; ; \\
'G':   & M^{'2}= - I \; ,\;\; & N^{'2} = - I  \; , & \;\; M'\;  'N = - 'N M'\;
\end{array} \right .
\eqno(2.8)
 $$

\noindent  and
 $$
 F\; S(k,
\bar{k}^{*}) = S(\bar{k}^{*},k) \; F\; , \;\; F \in \{ M , M', N, 'N \}  \; .
\eqno(2.9)
 $$

One can notice that the tables for the groups  $G_{M}$   and  $G_{N}$
happen to coincide; as well those of $G'$  and $'G$. This implies  that
the groups $G_{M}$ and $G_{N}$ represent the same abstract group, as well as    $G $ and $'G$.
Indeed, it is readily verified that
 $G_{M}$   and  $G_{N}$,
as well as $G'$   and $'G$, can be  inverted into each other by a similarity  transformation:
 $$
G_{N} = A \; G_{M} \; A^{-1} \; :     \;\; \;
 A \; S(k, \bar{k}^{*}) = S(\bar{k}^{*},k) \; A \; ,
 \eqno(2.10a)
$$
$$
 A\; M\; A^{-1} = + N \;,\; A \;M'\;  A^{-1} = + 'N \; ,
 $$
$$
A = const \left ( \begin{array}{cc} - i I & 0 \\ 0 & + I
\end{array} \right ) \; ;
$$
$$
 'G = A \;G'\; A^{-1}\; : \;\; A\;
S(k, \bar{k}^{*}) = S(\bar{k}^{*},k)\; A\; ,
\eqno(2.10b)
$$
$$
A\; M' \; A^{-1} = + 'N \; ,\;  A\; N \; A^{-1} = - M\; , \;
$$
$$
A = const  \left ( \begin{array}{cc} - i I & 0 \\ 0 & + I
\end{array} \right ) \;    .
$$

\noindent In other words,
above only four different covering groups were defined.
As long as   in literature all six variants are encountered,  all six  will be traced  below.

{\bf 3. Representations of extended spinor groups }

Now let us construct exact linear representations of the  groups
 $G_{M}, \;G_{N},\; G', \; 'G ,\; G , \; 'G'$.
It suffices   to consider  in detail only one  group, for definiteness let it be
$G_{M}$. Its multiplication table is
 $$
M^{2} = - I \; ,\;  M^{'2} = - I \; , \; M \; M'  = M\; M\; б
\eqno(3.1)
$$
$$
F\; S(k, \bar{k}^{*}) = S(\bar{k}^{*}, k)\; F \;
, \;\; ( \;\; F = M\; , M'\;\; ) \; ,
$$
$$
 (k_{1},
\bar{k}^{*}_{1} ) (k_{2}, \bar{k}^{*}_{2} ) = (<k_{1},k_{2}>,\;
<\bar{k}^{*}_{1} ,\bar{k}^{*}_{2}  >) \;.
$$

\noindent Here the symbol $< \;,\; >$  stands for the known
multiplication rule for  the vector-parameter on the group $SL(2.C)$:
$$
<k_{1}, \; k_{2}> =  (\; k^{0}_{1}\; k^{0}_{2}\; + \; \vec{k}_{1}\;
\vec{k}_{2} \; ; \;\; \vec{k}_{1} \; k^{0}_{2} \; + \; k^{0}_{2} \;
\vec{k}_{1} \; + \;  i \; [\; \vec{k}_{1} \; \vec{k}_{2} \;] \; ) \; .
$$

\noindent Let us look for solution of the problem $g \longrightarrow T(g)$  in the form
$$
T(g) = f(g)\;
g\; , \;\;\; g \in  G_{M} \; ,\;\;  f(g_{1}) \; f(g_{2}) = f(g_{1}
\;g_{2})    \;
\eqno(3.2a)
$$

\noindent where $f(g)$ is a numerical function on the group $G_{M}$.
Substituting  (3.2a)  into  (3.1) we get to
$$
[ f(M) ]^{2}= f(I) \; , \;  [ f(M') ]^{2}= f(-I) \; , \;\;
$$
$$
f( S(k,\bar{k}^{*}) ) = f( S(\bar{k}^{*},k) )\; , \;
$$
$$
f(S(k_{1}, \vec{k}^{*}_{1} ) )\; f( S(k_2), \vec{k}^{*}_{2} ) )
= f ( S (<k_{1}, k_{2}> \; , \; <\vec{k}^{*}_{1} \; , \;
\vec{k}^{*}_{2} >) )\; .
$$

\noindent There exist four different functions $f_{i}$ satisfying the equation above:
$$
\left.  \begin{array}{ccccc}
G_{M} & f_{1}(g) = & f_{2}(g) = & f_{3}(g) = & f_{4}(g) = \\[2mm]
S(k,\bar{k}^{*})  & + 1   & + 1   & + 1  & + 1  \\
M  & +  1 & - 1 & + 1 & - 1 \\
M'  & + 1 & - 1 & - 1 & + 1 \; .
\end{array} \right.
\eqno(3.2b)
$$

\noindent Correspondingly we have four representations $T_{i}(g)$
of the group  $G_{M}$.  In the same manner one can construct
analogous representation  $T_{i}(g) $ of remaining  five groups.
All those  can be described by the following table
$$
\left.  \begin{array}{cccccc}
& g = & T_{1}(g) = & T_{2}(g) = & T_{3}(g) = & T_{4}(g) = \\ [3mm]
   &  S(k,\bar{k}^{*}) &  S(k,\bar{k}^{*}) &
S(k,\bar{k}^{*}) &  S(k,\bar{k}^{*})  & S(k,\bar{k}^{*}) \\ [2mm]
 G_{M} & M & + M & - M & + M & - M \\
& M'  & + M' & - M' & -  M'  & + M'\\ [2mm]
G_{N} & N  & + N & - N & + N & - N \\
& 'N & +'N & -'N & -'N & +'N \\       [2mm]
G'  & M'  & + M'  & - M'  & + M '  & - M'  \\
& N & + N & - N &  - N & + N \\       [2mm]
'G & 'N & +' N & -'N & +' N & -'N \\
& M & + M & - M & - M & + M \\        [2mm]
G & M & + M & - M & + M & - M \\
& N & + N & - N & + N & - N \\        [2mm]
'G' & M' & +M' & - M' & + M'  & - M' \\
& 'N & +'N & -' N &  -' N & +' N
\end{array} \right.
\eqno(3.3)
 $$

For each  of the groups under consideration  one can ask a  question:
are the four representations $T_{i}(g)$ equivalent or not.
With the help of relations
$$
F = const  \left (
\begin{array}{cc} - I &  0  \\ 0 & + I \end{array} \right )
\;,\;\; F \; S(k,\bar{k}^{*}) \; F^{-1} = S(k,\bar{k}^{*}) \; ,
\eqno(3.4a)
$$
$$
 F \; M\;  F^{-1} = - M \; , \; F \;  M'\; F^{-1}
= - M'\; F\; , \; $$ $$ N \; F^{-1} = - N \;, \; F\; 'N \;F^{-1} =
- 'N    \;  $$

\noindent it is easily follows that  the type
$T_{2}(g)$ is equvalent to the type  $T_{1}(g)$, аs well as
$T_{4}(g)$ is equivalent to  $T_{3}(g)$:
$$
 T_{2}(g) = F
\; T_{1}(g) \;F^{-1} \;,\; T_{4}(g) = F \; T_{3}(g)\; F^{-1} \; .
\eqno(3.4b)
 $$

\begin{quotation}

Summarizing, we have got to the following:
for every of six groups only two non-equivalent representations
$g\; \rightarrow \; T(g) = f(g)\; g$  are possible:
$
T_{1}(g)  \sim T_{2}(g)  \;\;  и \;\; T_{3}(g)
\sim T_{4}(g) \; .
$
Evidently, this result does not depend on an explicit realization of
the discrete spinor transformations.
The above study of  the exact linear representations of the extended spinor groups
leads to a new concept of a space-time intrinsic parity of a fermion. In group-theoretical terms
$P$-parity and $T$-parity do not have any sense, instead only  their joint characteristic,
that might be called  $(PT)$-parity,  can be defined in the  group-theory
framework\footnote{It should be added  that
a very similar fact was noted  by  Yu. Shirokov [28] and De Vitt [32].}.

\end{quotation}

{\bf 4.
Representations of the coverings for $L^{\uparrow}_{+-}$ and
$L^{\uparrow \downarrow}_{+}$  }

Now we are going to consider the problem of linear representations of the spinor groups that
are supposedly cover the partly extended Lorentz groups
$L_{+-}^{\uparrow}$    and  $L_{+}^{\uparrow\downarrow}$  (improper orthochronous and
 proper non-orthochronous respectively).
Such coverings of partly extended groups  can be constructed by adding any one matrix from
 $M , M', N , 'N$ .

It is  known that orthogonal groups $L_{+-}^{\uparrow}$ and  $L_{+}^{\uparrow\downarrow}$ have
vector representations of only two types. The case of the group  $L_{+-}^{\uparrow}$:
$$
T_{1} = T_{3} \;\; :     \;\; L \Longrightarrow  L = (sgn \;
L_{0}^{\;\;0} ) \; L   \; \; ,
\eqno(4.1a)
$$
 $$
  T_{2} = T_{4}
\;\; :         \;\; L \Longrightarrow  L = (\det L ) L = (\det L)
( sgn \; L_{0}^{\;\;0} )\; L   \; .
\eqno(4.1b)
$$

\noindent The case of the group  $L_{+}^{\uparrow \downarrow}$ looks as
$$
T_{1} = T_{4}\; : \;\; L \Longrightarrow  L = (\det
L ) ( sgn \; L_{0}^{\;\;0} ) \; L    \; ,
\eqno(4.2a)
$$ $$
T_{2}
= T_{3}  \; \;: \;\;
 \; \Longrightarrow \; L = (\det  L ) \; L =  (sgn \;L_{0}^{\;\;0}) \; L  \; .
\eqno(4.2b) $$

Now let us proceed to partly extended spinor groups, coverings for
$L_{+}^{\uparrow \downarrow}$  and $L_{+-}^{\uparrow}$.
With the use of one additional discrete operation one can determine   four extended spinor
groups:
$$
\tilde{SL}(2.C)_{M}  = \{\; S(k,\bar{k}^{*}) \; \oplus \; M \}  \;\; \; \mbox{and  so on } .
\eqno(4.3)
 $$

\noindent In so doing
 extended groups   $\tilde{SL}(2.C)_{M}\; ,\;  \tilde{SL}(2.C)_{N}$ turn out to be isomorphic.
 Analogously,   $\tilde{SL}(2.C)_{M'}$ is isomorphic to $\tilde{SL}(2.C)_{'N}$.
Thus, there exist only two different spinor groups, each of them covers both
 $L_{+}^{\uparrow \downarrow}$  and $L_{+-}^{\uparrow}$:
$$
\tilde{SL}(2.C)_{M} \; \sim \; \tilde{SL}(2.C)_{N}\;\;  ,   \;\;
\tilde{SL}(2.C)_{M'} \; \sim \;  \tilde{SL}(2.C)_{'N} \; .
$$

Now, we are to list simplest  representations of these groups. The result obtaned is as follows:
all  above representations $T_{i}(g)$ (see  Section 3)   at confining them to sub-groups
$SL(2.C)_{M(N)}$   and $SL(2.C)_{M',('N)}$   lead to  representations changing into each other by a symmilarity
transformation. In other words,  in fact  there exists only one representation of these
partly extended spinor groups. This may be understood as  impossibility to determine any group-theoretical
parity concept ($P$ or $T$) within the limits of partly extended spinor groups.

{\bf 5. On reducing spinor groups to a real form}

Till now we have counted all spinor groups
$ G_{M} \sim G_{N} \; , \; \; G' \sim 'G \; , \; \; G \; , \; \; 'G'  \; $
 among possible candidates to be covering  of the full Lorentz group
$L^{\uparrow \downarrow}_{+-}$
It is desirable to have formulated some  extra argument to choose only one spinor group as
a genuine (physical) covering.

Let us draw attention to the fact that the used bispinor matrix (continuous one) $$
\Phi  = \left ( \begin{array}{c}
\xi ^{\alpha }  \\   \eta_{\dot{\alpha}}   \end{array} \right )  \; ;
\;\; \Phi ' = S(k, \bar{k}') \; \Phi \; , \;
S = \left ( \begin{array}{cc}
B(k)  &  0   \\ 0 &  B(\bar{k}^{*})  \end{array} \right )
\eqno(5.1)
$$

\noindent can be taken to a real-valued
form. This implies that in the bispinor space a special basis  can be found where the bispinor wave function
$$
\Phi _{M}(x) = \varphi (x)  + i \xi (x) \;\; \;\; \varphi^{*} (x) = \varphi (x)  \; , \;
\xi^{*}(x) = \xi (x)
$$

\noindent  transforms under  $SL(2.C)$ group by means of real $(4\times4)$-matrices.
Therefore, the real 4-spinors  $\varphi(x)$ and  $\xi (x)$, constituents of complex-valued $\Phi _{M}(x)$,
 transform as independent irreducible representations.
In physical context of real Majorana fermions this  reads as a group-theoretical permission to exist.
But these arguments have been  based only on continuous $SL(2.C)$-transformations, the idea is
to extend these on discrete operations too.

So we must find the answer to the  question of which of the extended spinor groups can be reduced
to a real form?
With this end in mind let us write down the  bispinor matrix in the form that does not depend on an
accidental basis choice\footnote{Above we  employed the Weyl basis.}:
$$
S(k,\bar{k}^{*}) =   {1 \over 2} (k_{0} + k^{*}_{0})  \; + \;
 {1 \over 2}  (k_{0} - k^{*}_{0} ) \; \gamma ^{5} \; + \;
(k_{1} + k^{*}_{1}) \; \sigma ^{01} \;+\;
(k_{1} - k^{*}_{1})\; i\; \sigma ^{23} \;+\;
$$
$$
(k_{2} +  k^{*}_{2}) \sigma ^{02} \;+ \;
(k_{2} - k^{*}_{2}) \;i\; \sigma ^{31}  \; +
(k_{3} + k^{*}_{3}) \; \sigma ^{03} \; + \;
(k_{3} - k^{*}_{3}) \; i\;  \sigma ^{13} \;  .
\eqno(5.2)
$$

\noindent The  form (5.2) being taken in the spinor frame
$$
\gamma ^{a} = \left ( \begin{array}{cc}
0 & \bar{\sigma }^{a} \\   \sigma ^{a} & 0  \end{array} \right ) \; , \;
\gamma ^{5} = - i \; \gamma ^{0} \; \gamma ^{1} \; \gamma ^{2} \;
\gamma ^{3} =  \left ( \begin{array}{cc}
- I &  0  \\  0  & + I    \end{array} \right )   \;
\eqno(5.3)
$$

\noindent gives the used above. But the form (5.2) is invariant under any similarity transformations:
$$
\gamma ^{a'}  = A\; \gamma ^{a} \; A^{-1} \; , \;\;
\Phi ' = A \; \Phi \; , \; S'(k, \bar{k}^{*}) =  A \; S_{спин.}(k,\bar{k}^{*})
\; A^{-1}            \; .
$$

\noindent
So, it remains to write down all used discrete operations $M , M' N ,'N$  in terms of Dirac matrices:
$$
M = + \gamma ^{0} \; , \; M' = + i\; \gamma ^{0} \;, \;
N = + i \; \gamma ^{5} \; \gamma ^{0} \; ,\;
 'N = - \gamma ^{5} \; \gamma ^{0}     \;
\eqno(5.4)
$$

\noindent and to take into account that  any Majorana basis satisfy  the relations
$$
(\gamma ^{a}_{M})^{*} = - \gamma ^{a}_{M} \; , \;\;
(\gamma ^{5}_{M}) ^{*} = - \gamma ^{5}_{M} \; , \;\;
( \sigma ^{ab}_{M})^{*} = \sigma ^{ab}_{M}  \; .
\eqno(5.5)
$$

\noindent  Therefore, in such frames the used discrete operations obey
$$
S^{*} = S\; , \;\;  M^{*} = - M \; , \;\; (M')^{*} = + M' \; , \;\;
N^{*} = - N \; , \;\; ('N)^{*} = + 'N  \; .
\eqno(5.6)
$$

\noindent Thus, six spinor groups behave under complex conjugation as indicated below
$$
\left. \begin{array}{cccccc}
G_{M} & G_{N}  & G' & 'G & G & 'G' \\ [1mm]
S^{*} = S & S^{*} = S & S^{*} = S & S^{*} = S & S^{*} = S & S^{*} = S \\
M^{*} = - M & N^{*} = -N & M'^{*}= +M' & 'N^{*}=+'N & M^{*} =-M & M'^{*}=+M \\
(M')^{*} = + M' & 'N^{*} = +'N & N^{*} = -N & M^{*}=-M & N^{*} =-N &
'M^{*} =+'M
\end{array} \right.
$$

\noindent From this it follows that only the  group  $'G'$ can be
reduced to a real-valued form. Only this group allows for real-valued spinor representations, Majorana
fermions\footnote{This  variant coincides with the known in the  literature Racah group [7].}.

\begin{center}
{\bf 6.  Discussion}
\end{center}

In the present  paper, the  problem of fermion parity is considered on the base of
investigating possible single-valued representations of
spinor coverings of the extended Lorentz group. It is shown that
in the frame of this theory does not exist, as separate concepts,
P-parity and T-parity for a fermion, instead only some unified
concept of (PT)-parity can be determined in a group-theoretical
Apparently, physics with spinor group in its base differs from that based on the
orthogonal group $L^{\uparrow \downarrow }_{+-}$ and only experiment must decide
the problem once and for all. It is needless to say that the task cannot be solved without
a thorough theoretical analysis of possible experimental verifications in both orthogonal and spinor framework.
language.

\begin{center}
{\bf References }
\end{center}

\noindent
1.
Lipschitz R.
Bonn, 1886.

\noindent
2.
Darboux G.
// Bull. des Sciences Math. 1905, Tom 9, 2 s\'{e}rie
(Ce memoire est un r\'{e}sum\'{e} des le\c{c}ons que l'auteur a faites
\`{a} la Sorbonne en 1900 et 1904.).

\noindent
3.
Cartan E.
// Bull. Soc. Math. France. 1913, Tom 41, P. 53-96.

\noindent
4.
Weyl H.
// Math. Z. 1924, Bd 23, S. 271-309;  1925, Bd 24, S. 329-395.

\noindent
5.
Schreier O.
// Hamb. Abh.  Math. Sem. 1925, Bd. 4, S. 15-32;
// Hamb. Abh.  Math. Sem. 1926, Bd. 5, S. 236-244.

\noindent
6.
Majorana E.
// Nuovo Cim. 1937, Vol. 14, P. 171-184.

\noindent
7.
Racah G.
Sulla simmetria tra particelle e antiparticell.
// Nuovo Cim.  1937, Vol. 14, P. 322.

\noindent
8.
Cartan E.
// Actualit\'{e}s Sci. et Ind. 1938, n. 643, Hermann, Paris;
1938, n. 701, Hermann, Paris.

\noindent
9.
Yang C.N., Tiomno J.
// Phys. Rev.  1950. Vol. 79, n. 3, P. 495-498.

\noindent
10.
Jarkov G.F.
JETP, 1950, Vol. 20, No 6,  P. 492-496.

\noindent
11.
Caianiello E.R.  // Nuovo Cim. 1951, Vol. 7, P. 534;
1951, Vol. 8, P. 749; 1952, Vol. 9, P. 336.

\noindent
12.
Wick G.C.,  Wigner E.P., Wightman A.S.
// Phys. Rev. 1952,  Vol. 88, P. 101.

\noindent
13.
Shapiro I.C.
// JETP. 1952, Vol. 22, No 5, P. 524-538.

\noindent
14.
Shapiro I.C.
// UFN. 1954, Vol. 53, No 1, P. 7-68.

\noindent
15.
Watanabe S.
// Rev. Mod. Phys. 1955, Vol. 27, P. 26-39;
1955, Vol.  27,  P. 40-76.

\noindent
16.
Heine V.
// Phys. Rev.  1957,  Vol. 107, n. 2, P. 620-623.

\noindent
17.
Широков Ю.М.
// ЖЭТФ. 1957, Том 33, Вып. 4,  С. 861-872;
1957, Том 33, Вып. 5,  С. 1196-1207;
1957, Том 33, Вып. 5,  С. 1208-1214;
1958, Том 34, Вып. 3,  С. 717-724;
1959, Том 36, Вып. 3,  С. 879-888.

\noindent
18.
Feinberg G., Weinberg S.
// Nuovo Cim. 1959,  Vol. 14, n. 3, P. 571-592.

\noindent
19.
Shirokov Yu.M.
// Nucl. Phys. 1960, Vol. 15, P. 1-12;
1960, Vol. 15, P. 13-15.

\noindent
20.
Богуш А.А.,  Федоров Ф.И.
// Докл. АН БССР. 1961, Том 5, n. 8, С. 327-330.

\noindent
21.
Sudarshan E.C.G.
// J. Math. Phys. 1965, Vol. 6, n. 8, P. 1329-1331.

\noindent
22.
Macfarlane A.J.
// J. Math. Phys. 1962, Vol. 3, P. 1116-1129.

\noindent
23.
Wigner E.
In: Group theoretical concepts and methods in elementary particle
physics. Ed. G\"{u}rsey F.N.Y., 1964 , Lectures at the Istambul Summer School
of theoretical physics. 1962, P. 37-80.

\noindent
24.
Lee T.D.
// Phys. Today. 1966, Vol. 19, n. 3, P. 23-31.

\noindent
25.
Lee T.D., Wick G.C.
// Phys. Rev. 1966, Vol. 148, n. 4, P. 1385-1404.

\noindent
26.
Goldberg H.
// Nuovo Cim. A. 1969, Vol. 60,  n. 4,  P. 509-518.

\noindent
27.
Wigner E.
Symmetry  and reflections. Indiana University Press, Bloomington - London, 1970.

\noindent
28.
Широков Ю.М.
// ЭЧАЯ. 1972 Том 3, Вып 3, С. 606-649; 1973. Том 4, Вып. 1, С. 42-78.

\noindent
29.
Yang C.N.
// J. de Physique. 1982, Colloque C8, P. 439.

\noindent
30.
Altmann S.L.
Rotations, quaternions, and double groups.  1986.

\noindent
31.
B.C. De Vitt.
Динамическая теория групп и полей.
Москва, Наука, 1987.

\noindent
32.
Sharma C.S.
// Nuovo Cim. B. 1989,  Vol. 103, P. 431-434.

\noindent
33.
Recai Erdem.
 12 pages; {\bf hep-th/9809054 V2}

\end{document}